\documentclass[conference,a4paper]{IEEEtran}
\IEEEoverridecommandlockouts

\usepackage{pgfplots}
\pgfplotsset{compat=1.18}
\usepackage{graphicx}
\usepackage{array}
\usepackage{cite}
\usepackage{amsmath,amssymb,amsfonts}
\usepackage{algorithmic}
\usepackage{textcomp}
\usepackage{xcolor}
\usepackage{float}
\usepackage{tikz}
\usetikzlibrary{arrows.meta, positioning, 
shapes.geometric}
\usepackage[hidelinks]{hyperref}
\usepackage{eso-pic}

\begin{document}

\AddToShipoutPictureFG*{%
  \AtPageLowerLeft{%
    \put(35,35){%
      \parbox{\dimexpr\paperwidth-70bp\relax}{%
        \centering
        \fontsize{6}{7}\selectfont
        \copyright{} 2026 IEEE. Personal use of this material is permitted.
        Permission from IEEE must be obtained for all other uses, in any
        current or future media, including reprinting/republishing this
        material for advertising or promotional purposes, creating new
        collective works, for resale or redistribution to servers or lists,
        or reuse of any copyrighted component of this work in other works.
      }%
    }%
  }%
}

\title{Comparative Analysis of Machine Learning Classifiers for Lightweight IoT Intrusion Detection Using Feature Reduction\\
}
\author{
\IEEEauthorblockN{Shaker Nawasra}
\IEEEauthorblockA{
School of Information Security \& Applied Computing \\
Eastern Michigan University \\
Ypsilanti, MI, USA \\
snawasra@emich.edu
}
\and
\IEEEauthorblockN{Munther Abualkibash}
\IEEEauthorblockA{
School of Information Security \& Applied Computing \\
Eastern Michigan University \\
Ypsilanti, MI, USA \\
mabualki@emich.edu
}
}
\maketitle

\noindent
{\footnotesize
\textbf{DOI:}
\href{https://doi.org/10.1109/AIBThings71029.2026.11771241}
{10.1109/AIBThings71029.2026.11771241}
}

\vspace{0.4em}

\begin{abstract}
Machine learning-based intrusion detection systems can improve IoT security, but high-dimensional traffic features and imbalanced datasets may reduce model reliability and deployment feasibility in resource-constrained environments. This study presents a lightweight binary IoT intrusion detection framework using an approximately balanced CICIoT2023-derived dataset, feature reduction, and comparative machine learning evaluation. The original multi-class attack labels were converted into two classes: Benign and Malicious. To reduce class bias, malicious traffic was randomly undersampled to obtain an approximately balanced binary dataset. Information Gain Ratio and Gini Decrease were then used to rank traffic features and generate reduced feature subsets of 38, 30, 20, 10, 5, and 3 features. Five machine learning classifiers, including Random Forest(RF), Decision Tree(DT), Naive Bayes(NB), Logistic Regression(LR), and Support Vector Machine(SVM), were evaluated using 5-fold cross-validation. The results show that Random Forest achieved the strongest overall performance. In particular, the 20-feature subset achieved accuracy of 99.1\% and MCC of 0.982, matching the full-feature model while reducing feature dimensionality by approximately 47.4\%. The 10-feature subset also maintained strong performance with accuracy of 98.9\% and MCC of 0.979, reducing the feature space by approximately 73.7\%. These findings demonstrate that balanced binary classification and progressive feature reduction can preserve detection effectiveness while improving lightweight IDS feasibility.
\end{abstract}
\vspace{0.5em}

\section{Introduction}

\subsection{Background}

The Internet of Things (IoT) has experienced rapid
growth in recent years, enabling billions of interconnected
devices to exchange data and provide intelligent services
across various domains, including healthcare, smart
homes, industrial systems, transportation, and critical
infrastructure~\cite{Atzori2010IoTSurvey}. The increasing
adoption of IoT technologies has significantly improved
automation, connectivity, and operational efficiency.

However, the large-scale deployment of IoT devices has
also introduced major cybersecurity challenges. IoT
networks are frequently targeted by cyberattacks such
as Distributed Denial of Service (DDoS), malware
infections, botnets, reconnaissance attacks, and data
manipulation attacks~\cite{SICARI2015146}. The
heterogeneous and highly connected nature of IoT
environments increases the attack surface and creates
additional security vulnerabilities.

Intrusion Detection Systems (IDSs) play an important
role in protecting IoT environments by monitoring
network traffic and identifying malicious activities.
Machine learning-based IDS approaches have gained
significant attention because they can automatically
detect abnormal traffic patterns and previously unseen
attacks with high detection capability~\cite{7307098}.

Despite these advantages, deploying machine learning-
based IDS frameworks in IoT environments remains
challenging due to the limited computational resources
available on many IoT devices. Most IoT systems operate
with constrained memory, storage capacity, processing
power, and energy resources~\cite{ALABA201710}.
As a result, lightweight intrusion detection solutions are
required to balance security effectiveness and resource
efficiency.

Feature reduction techniques provide a promising
approach for improving lightweight IDS feasibility by
reducing the number of input traffic attributes required
for classification. By minimizing dataset dimensionality,
machine learning models can achieve lower computational
complexity while maintaining acceptable intrusion
detection performance.

\subsection{Motivation}

Many existing machine learning-based IoT intrusion detection systems rely on multi-class classification using dozens of attack labels, which increases computational complexity, memory usage, training time, and inference overhead. These challenges reduce the feasibility of deploying IDS models on resource-constrained IoT and edge devices with limited processing power and storage capacity.

To address this issue, this study redesigns the intrusion detection process as a reduced weight binary classification problem by grouping all attack categories into a single Malicious class while preserving normal traffic as the Benign class. Instead of training the model using 34 different attack labels, the proposed framework focuses only on distinguishing between benign and malicious traffic. This significantly simplifies both the training and prediction processes while improving lightweight deployment suitability for IoT environments.

\begin{equation}
\operatorname{BinaryLabel}(x)=
\begin{cases}
\text{Benign}, & \text{if } x=\text{BENIGN},\\
\text{Malicious}, & \text{otherwise}.
\end{cases}
\label{eq:binarylabel}
\end{equation}

\subsection{Contributions}

The main contributions of this study are summarized
as follows:

\begin{itemize}

\item An approximately balanced binary IoT intrusion detection dataset was developed from the CICIoT2023 dataset by converting multi-class attack labels into two traffic classes, Benign and Malicious, and reducing majority-class bias through random undersampling.

\item Five lightweight machine learning classifiers, including Random Forest, Decision Tree, Naive Bayes, Logistic Regression, and Support Vector Machine, were comparatively evaluated for binary IoT intrusion detection.

\item The impact of progressive feature reduction was investigated using feature subsets of 38, 30, 20, 10, 5, and 3 features ranked by Information Gain Ratio and Gini Decrease.

\item The study identified a core group of network traffic features that consistently ranked highly across multiple feature importance methods. These features enabled substantial dimensionality reduction while preserving detection effectiveness, providing practical guidance for lightweight IoT IDS design.

\end{itemize}

\section{Related Work}

\subsection{Machine Learning for IoT IDS}

Machine learning has been widely used in IoT intrusion detection systems (IDSs) because of its ability to detect malicious network traffic automatically. Common classifiers include Random Forest, Decision Tree, Logistic Regression, Naive Bayes, and Support Vector Machine (SVM), which provide different tradeoffs between detection performance and computational efficiency.

Recent studies using the CICIoT2023 dataset have demonstrated effective detection of IoT attacks such as DDoS, DoS, spoofing, reconnaissance, web-based attacks, and Mirai malware attacks~\cite{s23135941}. Other studies have investigated scalable, deep-learning, and multi-class intrusion detection approaches for IoT environments~\cite{Hossain2025,fi17060230,Alrefaei2024,Alaiz_Moreton_2019}.

More recent work has specifically investigated feature selection and dimensionality reduction for lightweight IoT intrusion detection. Wakili and Bakkali used hybrid feature selection combining statistical, recursive, and tree-based methods~\cite{wakili2025resilient}. Hajjouz and Avksentieva applied hierarchical feature selection with Spearman correlation and CatBoost~\cite{hajjouz2024hierarchical}. Almahaqeri et al. used gain-based feature selection to reduce the CICIoT2023 feature space while lowering training and inference costs~\cite{almahaqeri2026optimized}. Kuo and Wu employed Mutual Information-based feature selection within a lightweight two-stage IDS designed for edge environments~\cite{kuo2026lightweight}.

To improve lightweight IDS feasibility, this study applies feature reduction and simplifies intrusion detection into two traffic classes: Benign and Malicious.

\subsection{Gap Analysis}

Existing studies demonstrate that feature selection can reduce IDS complexity while maintaining strong detection performance. However, recent CICIoT2023 approaches commonly focus on optimized ensemble, boosting, hierarchical, or two-stage architectures, often using multi-class attack classification, rather than systematically evaluating progressive feature reduction across multiple classical machine learning classifiers~\cite{wakili2025resilient,hajjouz2024hierarchical,almahaqeri2026optimized,kuo2026lightweight}.

To address this gap, this study simplifies the original attack labels into two classes, Benign and Malicious, and evaluates Random Forest, Decision Tree, Logistic Regression, Naive Bayes, and SVM using progressively reduced feature subsets of 38, 30, 20, 10, 5, and 3 features. In addition to accuracy, Matthews Correlation Coefficient (MCC) is emphasized because it considers all four confusion-matrix outcomes and provides a more balanced assessment of binary classification performance, particularly when class distributions are imbalanced. The objective is to identify the best balance between detection effectiveness, dimensionality reduction, and lightweight IoT deployment feasibility.

\section{Methodology}

\subsection{Proposed Machine Learning Framework}

Fig. 1 presents the workflow of the proposed lightweight IoT intrusion detection framework. The process begins with dataset preprocessing, including missing value imputation and feature normalization. 

The traffic labels are then converted into two main classes: Benign and Malicious to simplify the intrusion detection process and reduce classification complexity.

After preprocessing, the \textit{Number} feature was removed because it represents dataset metadata rather than actual network traffic behavior. Including this attribute could introduce data leakage and lead to misleading classification results by allowing the model to learn dataset-specific patterns instead of meaningful intrusion-related characteristics. 
Feature ranking is then performed using Information Gain Ratio and Gini Decrease to identify the most informative traffic attributes. Multiple reduced feature subsets are generated and evaluated to analyze the tradeoff between lightweight feasibility and detection performance.

Several machine learning classifiers, including Random Forest, Decision Tree, Naive Bayes, Logistic Regression, and Support Vector Machine, are trained and evaluated using 5-fold cross-validation.

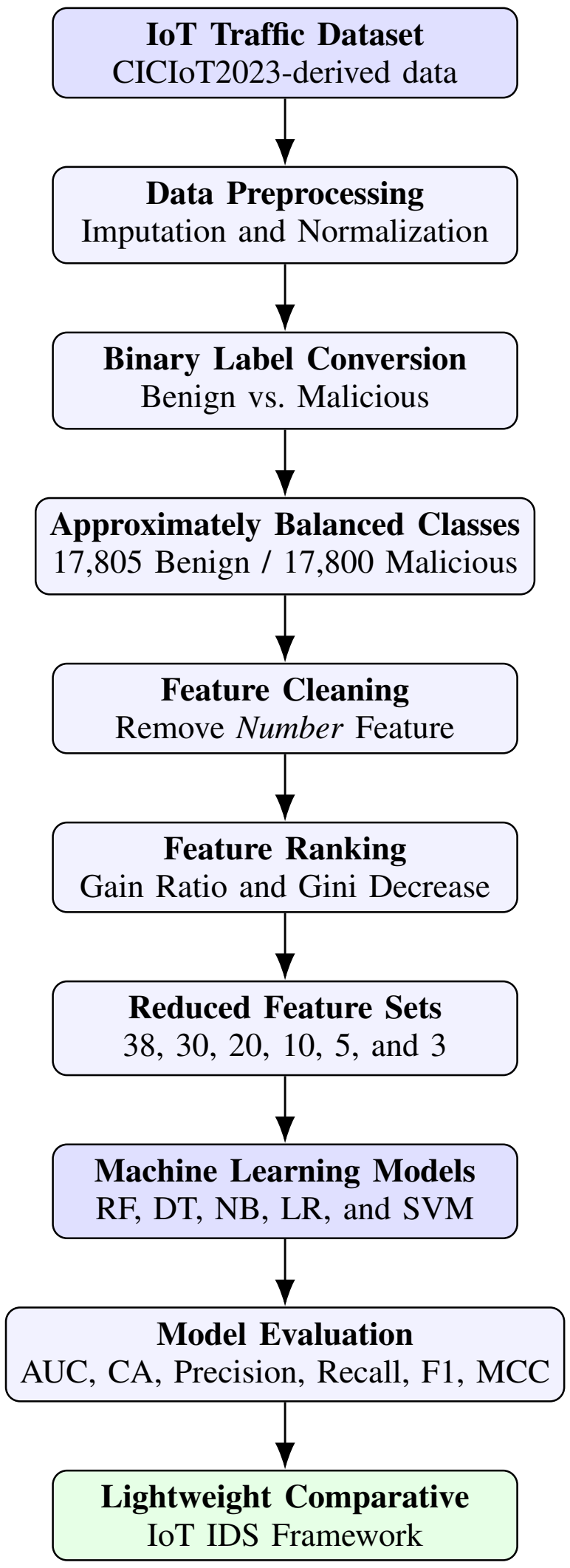
\begin{figure}[!h]
\centering
\resizebox{0.6\columnwidth}{!}{
\begin{tikzpicture}[
    node distance=0.50cm,
    box/.style={
        rectangle,
        rounded corners=3pt,
        draw=black,
        line width=0.4pt,
        align=center,
        minimum width=3.45cm,
        minimum height=0.65cm,
        inner sep=3pt,
        font=\scriptsize,
        fill=blue!5
    },
    important/.style={
        box,
        fill=blue!12
    },
    finalbox/.style={
        box,
        fill=green!10
    },
    arrow/.style={-{Latex[length=2mm]}, line width=0.45pt}
]

\node[important] (data) {\textbf{IoT Traffic Dataset}\\CICIoT2023-derived data};

\node[box, below=of data] (prep) {\textbf{Data Preprocessing}\\Imputation and Normalization};

\node[box, below=of prep] (binary) {\textbf{Binary Label Conversion}\\Benign vs. Malicious};

\node[box, below=of binary] (balance) {\textbf{Approximately Balanced Classes}\\17,805 Benign / 17,800 Malicious};

\node[box, below=of balance] (clean) {\textbf{Feature Cleaning}\\Remove \textit{Number} Feature};

\node[box, below=of clean] (rank) {\textbf{Feature Ranking}\\Gain Ratio and Gini Decrease};

\node[box, below=of rank] (subset) {\textbf{Reduced Feature Sets}\\38, 30, 20, 10, 5, and 3};

\node[important, below=of subset] (models) {\textbf{Machine Learning Models}\\RF, DT, NB, LR, and SVM};

\node[box, below=of models] (eval)
{\textbf{Model Evaluation}\\AUC, CA, Precision, Recall, F1, MCC};

\node[finalbox, below=of eval] (final)
{\textbf{Lightweight Comparative}\\IoT IDS Framework};

\draw[arrow] (data) -- (prep);
\draw[arrow] (prep) -- (binary);
\draw[arrow] (binary) -- (balance);
\draw[arrow] (balance) -- (clean);
\draw[arrow] (clean) -- (rank);
\draw[arrow] (rank) -- (subset);
\draw[arrow] (subset) -- (models);
\draw[arrow] (models) -- (eval);
\draw[arrow] (eval) -- (final);

\end{tikzpicture}
}
\caption{Proposed lightweight IoT intrusion detection framework integrating binary label conversion, class balancing, feature reduction, and comparative machine learning evaluation.}
\label{fig:framework}
\end{figure}
\subsection{Dataset and Preprocessing}
The experiments were conducted using Merged24 from the CICIoT2023 dataset. The original dataset contained 743,201 traffic records and 40 attributes. The original attack labels were converted into a binary classification problem by labeling normal traffic as Benign and grouping all attack categories into a single Malicious class. Because the original dataset was highly imbalanced, with approximately 97.6\% malicious traffic and 2.4\% benign traffic, random undersampling was applied to create an approximately balanced dataset. All benign records were retained, and an approximately equal number of malicious records were randomly selected. The resulting balanced dataset contained 35,605 records (17,805 benign and 17,800 malicious samples) and was used for feature ranking, classifier training, and evaluation.

Data preprocessing was performed using Orange Data Mining~\cite{demsar2013orange}. Missing values were handled using imputation, and numeric features were normalized to support fair comparison across classifiers, especially distance-based and optimization-based models such as SVM and Logistic Regression. The Number attribute was removed because it does not represent meaningful network behavior and may introduce data leakage. The final dataset contained 38 traffic features and one binary target variable.

\begin{table}[!h]
\caption{Dataset and Preprocessing Summary}
\centering
\resizebox{\columnwidth}{!}{
\begin{tabular}{|l|l|}
\hline
\textbf{Property} & \textbf{Description} \\ \hline
Original Dataset Size & 743,201 records \\ \hline
Original Attributes & 40 \\ \hline
Traffic Classes & Benign, Malicious \\ \hline
Original Class Distribution & 97.6\% Malicious, 2.4\% Benign \\ \hline
Approximately Balanced Dataset& 35,605 records \\ \hline
Removed Feature & \textit{Number} \\ \hline
Final Features & 38 \\ \hline
Preprocessing & Imputation and Normalization \\ \hline
Platform & Orange Data Mining \\ \hline
\end{tabular}
}
\label{tab:dataset}
\end{table}

\subsection{Feature Reduction and Classification}
Feature ranking was performed using Information Gain Ratio~\cite{quinlan1993c45} and Gini Decrease~\cite{breiman1984classification} to identify the most informative network traffic attributes for intrusion detection.
Information Gain Ratio measures how effectively a feature reduces class uncertainty while minimizing bias toward features with many distinct values. Gini Decrease evaluates feature importance based on impurity reduction during decision tree splitting. These methods were selected because they provide efficient and lightweight feature importance evaluation suitable for IoT intrusion detection environments. Multiple reduced feature subsets consisting of 38, 30, 20, 10, 5, and 3 features were generated to analyze the tradeoff between lightweight feasibility and detection performance.

Several machine learning classifiers, including Random Forest, Decision Tree, Naive Bayes, Logistic Regression, and Support Vector Machine, were trained and evaluated using the reduced feature subsets.
Feature reduction contributes directly to lightweight IDS deployment by decreasing the amount of traffic information that must be collected, stored, processed, and analyzed during both training and inference. Smaller feature sets reduce memory consumption, computational complexity, feature extraction overhead, and communication costs in distributed IoT environments. These benefits are particularly important for IoT gateways, edge devices, and embedded systems that operate under strict resource constraints. Consequently, evaluating the impact of progressive feature reduction provides insight into the tradeoff between detection effectiveness and deployment feasibility.

\subsection{Experimental Setup}
The experiments were conducted using the Orange Data Mining platform. Five machine learning classifiers, namely Random Forest (RF), Decision Tree (DT), Logistic Regression (LR), Naive Bayes (NB), and Support Vector Machine (SVM), were evaluated using 5-fold cross-validation to ensure consistent and reliable performance estimation. To analyze the impact of feature reduction, experiments were performed using feature subsets containing 38, 30, 20, 10, 5, and 3 features derived from Information Gain Ratio and Gini Decrease rankings. Model performance was assessed using Accuracy (CA), Precision, Recall, F1-score, Area Under the ROC Curve (AUC), and Matthews Correlation Coefficient (MCC). Particular emphasis was placed on MCC because it provides a balanced evaluation of binary classification performance and remains reliable when class distributions are imbalanced.

\section{Results and Discussion}
\subsection{Impact of Feature Reduction on Classifier Performance}

The performance of Random Forest (RF), Logistic Regression (LR),
Decision Tree (DT), Naive Bayes (NB), and Support Vector Machine
(SVM) was evaluated across six feature subsets using 5-fold
cross-validation. Table~II summarizes the MCC results obtained by
each classifier. MCC was selected as the primary evaluation metric
because it considers true positives, true negatives, false positives,
and false negatives simultaneously.

\begin{table}[!h]
\centering
\caption{Comparative MCC Performance Across Feature Subsets}
\label{tab:mcc_comparison}
\resizebox{\columnwidth}{!}{
\begin{tabular}{ccccccc}
\hline
\textbf{Features} &
\textbf{RF} &
\textbf{LR} &
\textbf{DT} &
\textbf{NB} &
\textbf{SVM} &
\textbf{Best Model} \\
\hline
38 & 0.982 & 0.976 & 0.960 & 0.937 & 0.841 & RF \\
30 & 0.982 & 0.976 & 0.960 & 0.940 & 0.690 & RF \\
20 & 0.982 & 0.975 & 0.960 & 0.940 & 0.690 & RF \\
10 & 0.979 & 0.942 & 0.953 & 0.885 & 0.690 & RF \\
5  & 0.935 & 0.864 & 0.931 & 0.870 & 0.029 & RF \\
3  & 0.935 & 0.864 & 0.931 & 0.870 & 0.029 & RF \\
\hline
\end{tabular}
}
\end{table}

The results show that Random Forest consistently achieved the
highest MCC across all feature subsets. RF maintained an MCC of
0.982 when the feature set was reduced from 38 to 20 features,
indicating that substantial dimensionality reduction was possible
without measurable loss in detection effectiveness. Logistic
Regression achieved the second-best overall performance, while
Decision Tree remained stable, maintaining an MCC above 0.93 even
with only three features. Naive Bayes produced reasonable but lower
performance than RF, LR, and DT.

SVM was the most sensitive to feature reduction. Its MCC decreased
from 0.841 with 38 features to 0.029 with five and three features,
indicating reduced robustness under aggressive dimensionality
reduction. Overall, Random Forest provided the strongest balance
between detection performance and feature reduction, making it the
most suitable classifier for the lightweight IoT IDS considered in
this study.

\subsection{Confusion Matrix Analysis}
To further evaluate the effectiveness of the proposed framework, Fig. 2 presents the confusion matrix obtained by the Random Forest classifier using the 20-feature subset. The matrix aggregates the out-of-fold predictions generated during 5-fold cross-validation. Thus, each of the 35,605 dataset instances contributes one prediction to the final confusion matrix. This configuration was selected because it achieved the same overall detection performance as the full 38-feature model while substantially reducing the feature space by approximately 47.4\%.

The confusion matrix shows that the classifier correctly
identified 17,785 benign traffic instances and 17,504
malicious traffic instances. 

Only 20 benign samples were
incorrectly classified as malicious, while 296 malicious
samples were incorrectly classified as benign.

The very small number of false positives demonstrates
the ability of the proposed framework to minimize false
alarm generation, which is particularly important in
practical intrusion detection environments where
excessive alerts can overwhelm security analysts and
reduce operational efficiency. Furthermore, the low
number of false negatives indicates that most malicious
traffic instances were successfully detected.

These findings are consistent with the high classification
metrics reported in Tables~\ref{tab:mcc_comparison} and
\ref{tab:feature_reduction}, where the Random Forest
classifier achieved an MCC of 0.982 and an accuracy of
99.1\%. Overall, the confusion matrix confirms that the
20-feature Random Forest model provides an effective
balance between lightweight deployment requirements
and reliable intrusion detection performance.

\begin{figure}[!h]
\centering
\includegraphics[width=0.83\columnwidth]{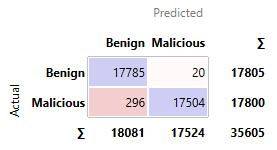}
\caption{Confusion matrix of the Random Forest classifier using the 20-feature subset.}
\label{fig:cm_rf20}
\end{figure}

Only 20 benign traffic instances out of 17,805 were incorrectly classified as malicious, corresponding to a false positive rate of approximately 0.11\%.

\subsection{Feature Reduction Analysis}

Table~\ref{tab:feature_reduction} summarizes the best
classification performance achieved across the evaluated
feature subsets. The experiments investigated the impact
of progressive feature reduction on lightweight IoT
intrusion detection using approximately balanced binary dataset and five machine learning classifiers.

The results indicate that Random Forest consistently
achieved the highest overall performance across all
feature subsets. Using the complete 38-feature set
produced an accuracy of 99.1\% and an MCC of 0.982.
Interestingly, reducing the feature space to 30 and
20 features preserved exactly the same detection
performance while eliminating 21.1\% and 47.4\% of
the original features, respectively.

The 10-feature subset also maintained strong detection
capability, achieving 98.9\% accuracy and an MCC of
0.979 while reducing feature dimensionality by 73.7\%.
These findings demonstrate that a large portion of the
original feature space contributes little additional
discriminative information for binary intrusion detection.

Although the 5-feature and 3-feature subsets achieved
reasonable performance, their MCC values decreased to
0.935, indicating a reduction in classification stability
under aggressive feature reduction. Therefore, excessive
dimensionality reduction may remove useful traffic
characteristics required for reliable intrusion detection.

From a lightweight deployment perspective, reducing
the feature space decreases the amount of data that must
be processed during both training and inference. Smaller
feature subsets reduce memory requirements, storage
overhead, and computational complexity, making the
resulting IDS more suitable for IoT gateways, edge
devices, and other resource-constrained environments.

Overall, the 20-feature configuration achieved the best
tradeoff between detection effectiveness and feature
reduction. This subset preserved the same accuracy,
F1-score, and MCC as the full-feature model while
eliminating nearly half of the original feature space.

\begin{table}[!h]
\centering
\caption{Best Detection Performance Across Feature Subsets}
\label{tab:feature_reduction}
\scriptsize
\begin{tabular}{cccccc}
\hline
\textbf{Feat.} &
\textbf{Model} &
\textbf{Acc.} &
\textbf{F1} &
\textbf{MCC} &
\textbf{Feature Reduction} \\
\hline
38 & RF & 0.991 & 0.991 & 0.982 & 0.0\% \\
30 & RF & 0.991 & 0.991 & 0.982 & 21.1\% \\
20 & RF & 0.991 & 0.991 & 0.982 & 47.4\% \\
10 & RF & 0.989 & 0.989 & 0.979 & 73.7\% \\
5  & RF & 0.968 & 0.968 & 0.935 & 86.8\% \\
3  & RF & 0.968 & 0.968 & 0.935 & 92.1\% \\
\hline
\end{tabular}
\end{table}

\section{Limitations}

Although the proposed framework achieved strong binary intrusion detection performance, several limitations should be acknowledged.

First, the experiments were conducted using a single subset (Merged24) of the CICIoT2023 dataset. While the dataset contains a wide range of IoT attack scenarios, evaluating the proposed framework on additional IoT intrusion detection datasets would provide a more comprehensive assessment of model generalization and robustness across diverse network environments.

Second, the framework was evaluated in an offline experimental setting using the Orange Data Mining platform. Consequently, practical deployment characteristics such as inference latency, memory consumption, CPU utilization, energy requirements, and throughput were not directly measured.

Third, the study employed random undersampling to obtain an approximately balanced dataset and reduce majority-class bias. While this approach enabled fair classifier comparison, it may not fully represent real-world IoT networks where malicious and benign traffic distributions are often highly imbalanced. Future investigations should examine the performance of the proposed framework under realistic traffic distributions.

Despite these limitations, the experimental results demonstrate that substantial feature reduction can be achieved while maintaining strong intrusion detection performance, supporting the feasibility of lightweight machine learning-based IDS solutions for resource-constrained IoT environments.

\section{Conclusion}

This study presented a lightweight IoT intrusion detection framework based on binary classification, class balancing, feature reduction, and comparative machine learning evaluation. By transforming the original multi-class CICIoT2023 dataset into a binary classification problem, the proposed approach simplified intrusion detection while maintaining strong detection performance.

Experimental results showed that Random Forest consistently outperformed the other evaluated classifiers across all feature subsets. Notably, the 20-feature configuration achieved the same MCC (0.982) and accuracy (99.1\%) as the full-feature model while reducing the feature space by approximately 47.4\%. The 10-feature subset also maintained strong performance while reducing dimensionality by 73.7\%.

The findings demonstrate that substantial feature reduction can be achieved without significant degradation in detection effectiveness. Furthermore, removing the \textit{Number} attribute improved model reliability by eliminating a potential source of data leakage. Overall, the proposed framework provides an effective and computationally efficient approach for lightweight IoT intrusion detection in resource-constrained environments.

\section{Future Work}

Future work will focus on evaluating the proposed framework using additional IoT intrusion detection datasets to further assess its generalization capability, robustness, and applicability across diverse network environments and attack scenarios. Cross-dataset validation will help determine whether the identified feature subsets remain effective beyond the CICIoT2023 dataset and can support reliable intrusion detection under varying traffic characteristics.

Another important direction is the deployment and evaluation of the proposed framework in real-time IoT and edge computing environments. Future experiments may investigate processing latency, memory consumption, computational overhead, and scalability when operating on resource-constrained devices such as IoT gateways, embedded systems, and edge nodes. Such evaluations would provide deeper insight into the practical feasibility of lightweight intrusion detection in operational environments.

Future studies may also explore alternative feature ranking and selection techniques, including wrapper-based and hybrid approaches, to identify compact feature subsets that further reduce dimensionality while maintaining strong detection performance. In addition, ensemble learning strategies and advanced machine learning algorithms may be investigated to improve classification robustness and resilience against evolving cyber threats.
\newpage
\bibliographystyle{IEEEtran}
\bibliography{references}

@book{breiman1984classification,
  title={Classification and Regression Trees},
  author={Breiman, Leo and Friedman, Jerome H. and Olshen, Richard A. and Stone, Charles J.},
  year={1984},
  edition={1st},
  publisher={Chapman and Hall/CRC},
  doi={10.1201/9781315139470},
    note    = {doi: 10.1201/9781315139470}
}

@article{ALABA201710,
title = {{Internet of Things} security: A survey},
journal = {Journal of Network and Computer Applications},
volume = {88},
pages = {10-28},
year = {2017},
issn = {1084-8045},
doi = {10.1016/j.jnca.2017.04.002},
author = {Fadele Ayotunde Alaba and Mazliza Othman and Ibrahim Abaker Targio Hashem and Faiz Alotaibi},
note    = {doi: 10.1016/j.jnca.2017.04.002}
}

@article{SICARI2015146,
title = {Security, privacy and trust in {Internet of Things}: The road ahead},
journal = {Computer Networks},
volume = {76},
pages = {146-164},
year = {2015},
issn = {1389-1286},
doi = {10.1016/j.comnet.2014.11.008},
author = {S. Sicari and A. Rizzardi and L.A. Grieco and A. Coen-Porisini},
note    = {doi: 10.1016/j.comnet.2014.11.008}
}

@ARTICLE{7307098,
  author={Buczak, Anna L. and Guven, Erhan},
  journal={IEEE Communications Surveys \& Tutorials}, 
  title={A Survey of Data Mining and Machine Learning Methods for Cyber Security Intrusion Detection}, 
  year={2016},
  volume={18},
  number={2},
  pages={1153-1176},
  doi={10.1109/COMST.2015.2494502},
  note    = {doi: 10.1109/COMST.2015.2494502}
  }

@article{Atzori2010IoTSurvey,
  author  = {Luigi Atzori and Antonio Iera and Giacomo Morabito},
  title   = {The {Internet of Things}: A Survey},
  journal = {Computer Networks},
  volume  = {54},
  number  = {15},
  pages   = {2787--2805},
  month   = oct,
  year    = {2010},
  doi     = {10.1016/j.comnet.2010.05.010},
  note    = {doi: 10.1016/j.comnet.2010.05.010}
}

@Article{s23135941,
  author = {Neto, Euclides Carlos Pinto and Dadkhah, Sajjad and Ferreira, Raphael and Zohourian, Alireza and Lu, Rongxing and Ghorbani, Ali A.},
  title = {{CICIoT2023}: A Real-Time Dataset and Benchmark for Large-Scale Attacks in {IoT} Environment},
  journal = {Sensors},
  volume = {23},
  number = {13},
  pages = {5941},
  year = {2023},
  doi = {10.3390/s23135941},
  note    = {doi: 10.3390/s23135941}
}

@article{Hossain2025,
  author    = {Md. Alamgir Hossain},
  title     = {Deep learning-based intrusion detection for {IoT} networks: a scalable and efficient approach},
  journal   = {EURASIP Journal on Information Security},
  year      = {2025},
  volume    = {2025},
  number    = {1},
  pages     = {28},
  doi       = {10.1186/s13635-025-00202-w},
 issn      = {2510-523X},
 publisher = {SpringerOpen},
 note    = {doi: 10.1186/s13635-025-00202-w}
}

@Article{fi17060230,
AUTHOR = {Seyedkolaei, Ali Abdi and Mahmoudi, Fatemeh and García, José},
TITLE = {A Deep Learning Approach for Multiclass Attack Classification in {IoT} and {IIoT} Networks Using Convolutional Neural Networks},
JOURNAL = {Future Internet},
VOLUME = {17},
YEAR = {2025},
NUMBER = {6},
ARTICLE-NUMBER = {230},
ISSN = {1999-5903},
DOI = {10.3390/fi17060230},
note    = {doi: 10.3390/fi17060230}
}

@article{Alrefaei2024,
  author    = {Ahmed Alrefaei and Mohammad Ilyas},
  title     = {Using Machine Learning Multiclass Classification Technique to Detect {IoT} Attacks in Real Time},
  journal   = {Sensors},
  year      = {2024},
  volume    = {24},
  number    = {14},
  pages     = {4516},
  doi       = {10.3390/s24144516},
  publisher = {MDPI},
  note    = {doi: 10.3390/s24144516}
}

@article{Alaiz_Moreton_2019,
   title={Multiclass Classification Procedure for Detecting Attacks on {MQTT‐IoT Protocol}},
   volume={2019},
   ISSN={1099-0526},
   DOI={10.1155/2019/6516253},
   number={1},
   journal={Complexity},
   publisher={Wiley},
   author={Alaiz-Moreton, Hector and Aveleira-Mata, Jose and Ondicol-Garcia, Jorge and Muñoz-Castañeda, Angel Luis and García, Isaías and Benavides, Carmen},
   editor={Krzemień, Alicja},
   year={2019},  
    month={Jan} ,
   note    = {doi: 10.1155/2019/6516253}
   }

@article{wakili2025resilient,
  author  = {Abubakar Wakili and Sara Bakkali},
  title   = {A Resilient {IoT} Intrusion Detection System Using Hybrid Feature Selection and Explainable Ensemble Learning},
  journal = {Results in Engineering},
  volume  = {28},
  pages   = {107392},
  year    = {2025},
  doi     = {10.1016/j.rineng.2025.107392},
  note    = {doi: 10.1016/j.rineng.2025.107392}
}

@article{hajjouz2024hierarchical,
  author  = {Abdulkader Hajjouz and Elena Avksentieva},
  title   = {Optimizing Intrusion Detection for {DoS}, {DDoS}, and {Mirai} Attack Subtypes Using Hierarchical Feature Selection and {CatBoost} on the {CICIoT2023} Dataset},
  journal = {Data and Metadata},
  volume  = {3},
  pages   = {577},
  year    = {2024},
  doi     = {10.56294/dm2024577},
  note    = {doi: 10.56294/dm2024577}
}

@article{almahaqeri2026optimized,
  author  = {Salah A. Almahaqeri and Mohammed Hashem Almourish and Adel A. Nasser and Abed Saif Ahmed Alghawli and Amani A. K. Elsayed and Ali N. Alhejoj},
  title   = {An Optimized Gradient Boosting Framework for {IoT} Intrusion Detection: A Comprehensive Evaluation on the {CICIoT2023} Dataset},
  journal = {Scientific Reports},
  volume  = {16},
  pages   = {16909},
  year    = {2026},
  doi     = {10.1038/s41598-026-47399-5} ,
  note    = {doi: 10.1038/s41598-026-47399-5}
}

@article{kuo2026lightweight,
  author  = {Chung-Wei Kuo and Cheng-Xuan Wu},
  title   = {A Lightweight Two-Stage Intrusion Detection Framework Optimized for Edge-Based {IoT} Environments},
  journal = {Computers, Materials \& Continua},
  volume  = {87},
  number  = {3},
  pages   = {53},
  year    = {2026},
  doi     = {10.32604/cmc.2026.076767},
  note    = {doi: 10.32604/cmc.2026.076767}
}

@article{demsar2013orange,
  author  = {Janez Dem{\v{s}}ar and Toma{\v{z}} Curk and Ale{\v{s}} Erjavec and {\v{C}}rt Gorup and Toma{\v{z}} Ho{\v{c}}evar and Mitar Milutinovi{\v{c}} and Martin Mo{\v{z}}ina and Matija Polajnar and Marko Toplak and An{\v{z}}e Stari{\v{c}} and Miha {\v{S}}tajdohar and Lan Umek and Lan {\v{Z}}agar and Jure {\v{Z}}bontar and Marinka {\v{Z}}itnik and Bla{\v{z}} Zupan},
  title   = {Orange: Data Mining Toolbox in {Python}},
  journal = {Journal of Machine Learning Research},
  volume  = {14},
  pages   = {2349--2353},
  year    = {2013},
  note    = {[Online]. Available: https://jmlr.org/papers/v14/demsar13a.html}
}

@book{quinlan1993c45,
  author    = {J. Ross Quinlan},
  title     = {{C4.5}: Programs for Machine Learning},
  publisher = {Morgan Kaufmann},
  year      = {1993},
  doi       = {10.1016/C2009-0-27846-9},
  note    = {doi: 10.1016/C2009-0-27846-9}
}
\end{document}